\documentclass[aps,pre,twocolumn]{revtex4}

\usepackage[dvips]{graphicx}
\usepackage{amssymb}

\begin{document}

\title{Engineering Fano resonances in discrete arrays}

\author{Andrey E. Miroshnichenko}
\author{Yuri S. Kivshar}

\affiliation{Nonlinear Physics Centre and Centre for Ultra-high
bandwidth Devices for Optical Systems (CUDOS), Research School of
Physical Sciences and Engineering, Australian National University,
Canberra ACT 0200, Australia}

\begin{abstract}
We study transmission properties of discrete arrays composed of a
linear waveguide coupled to a system of $N$ side defect states.
This simple system can be used to model discrete networks of
coupled defect modes in photonic crystals, complex waveguide
arrays in two-dimensional nonlinear lattices, and ring-resonator
structures. We demonstrate the basic principles of the resonant
scattering management through engineering Fano resonances and find
exact results for the wave transmission coefficient. We reveal
conditions for perfect reflections and transmissions due to either
destructive or constructive interferences, and associate them with
Fano resonances, also demonstrating how these resonances can be
tuned by nonlinear defects.
\end{abstract}

\maketitle

\section{Introduction}

During last decade we observe a growing interest in theoretical
and experimental studies of different types of resonant wave
phenomena associated with either direct or indirect manifestation
of the classical Fano resonance~\cite{uf61} in nanoscale devices
with side-coupled waveguiding
structures~\cite{goeres00}-\cite{vljpv04}. These structures can be
presented as one or more waveguides in which forward and backward
propagating waves are indirectly coupled to each other via one or
more mediating resonant cavities or defect states. The well-known
systems for realizing these structures are based on a straight
photonic-crystal waveguide with a number of side defect
modes~\cite{fan}, micro-ring resonator structures in which two
channel waveguides are side-coupled to micro-ring
resonators~\cite{ring} or variety of bend photonic-crystal waveguides \cite{aemysk05}.
Similar structures can be created in the
discrete networks extensively discussed for routing and switching
of discrete optical solitons~\cite{chris}.

In all such structures, the forward and backward propagating modes
within the waveguide are coupled via the defects; the transmission
becomes highly sensitive to the resonant properties of the defect
states, and it is usually associated with the so-called Fano
resonances. Indeed, the underlying physics of the Fano resonances
finds its origin in {\em wave interference} which occurs in the
systems characterized by one or several discrete energy states
that interact with the continuum spectrum. In the corresponding
transmission dependencies, the interference effect leads to either
{\em perfect transmission} or {\em perfect reflection}, producing
a sharp asymmetric response. This kind of the wave resonance is
also common in different interferometer devices such as the
Aharonov-Bohm interferometer~\cite{AB} and the Mach-Zehnder
interferometer~\cite{MZ}.

One of the simplest models that can be used to study the Fano
resonances in discrete networks as well as the resonant coupling
and interaction between discrete degrees of freedom and a
continuum spectrum is the so-called Fano-Anderson
model~\cite{mahan}. It describes a linear array of coupled
elements (e.g., effective particles or defect modes) with the
nearest-neighbor interaction coupled to one or several defect
states through a local coupling. Such a simple discrete model
allows to describe the basic physics of the Fano resonances in a
rather simple way including the nonlinear and bistability
regimes~\cite{aemsfmsfysk05}. In particular, this model allows to
derive analytical results for the wave transmission and
reflection, and it may serve as a guideline for the analysis of
more complicated physical models associated with the Fano
resonance.

In this paper, we study the transmission properties of discrete
networks composed of linear arrays of interacting elements coupled
to systems of $N$ side defects described by the generalized discrete
Fano-Anderson model. This model allows us to find exact solutions
for the wave transmission coefficient and the conditions for the
perfect reflections and transmissions due to either destructive or
constructive interference. Using these results, we demonstrate and
explain the basic principles of the resonant scattering under the
condition of the Fano resonances, and also suggest the concept of
the Fano resonance engineering. In particular, for several different
examples we demonstrate  that in the presence of a defect the
destructive wave interference remains always resonant, while the
constructive wave interference could be or could not be resonant. As
a result, this brings us to the general conclusion that the main
feature of the Fano resonance is the resonant reflection but not
transmission. We also demonstrate how the Fano resonances can be
tuned by introducing nonlinear defects into a discrete network.
The results are quite general and can be applied to different 
physical systems such as quantum dots or photonic crystal waveguides,
for example.

The paper is organized as follows. In Sec.~II we introduce our
discrete model describing a linear chain with $N$ defects and
describe the main features of the linear wave transmission. In
particular, we define the conditions for both resonant
transmission and reflection due to the interaction with a side
chain of $N$ defects coupled locally to the main array. In
Sec.~III we demonstrate how the Fano resonance in a linear system
can be tuned by introducing nonlinear defects. Finally, Sec. IV
concludes the paper.

\section{Linear transmission}

%
\begin{figure}[htb]
\centerline{
\includegraphics[width=80mm]{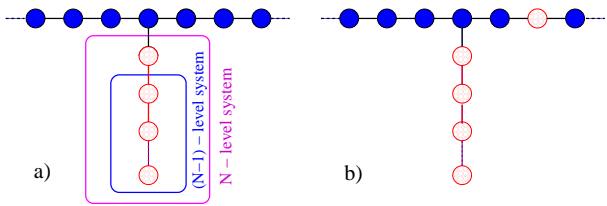}}
\caption{\label{fig:fig1}(Color online) Schematic view of the generalized
Fano-Anderson model with a locally coupled $N$-defect chain
without (a) or with (b) additional defects in the main array.
}%
\end{figure}

We consider the generalized linear Fano-Anderson model that
describes an infinite array of interacting elements (e.g. effective
particles) coupled locally to a complex side defect (the so-called
Fano defect) characterized by more than one degrees of freedom, as
shown schematically in a specific example of Fig.~\ref{fig:fig1}(a).
From the viewpoint of the Fano resonance, each degree of freedom of
the defect chain contributes with an additional local path for the
wave scattering, or, in other words, generates an additional
discrete state. Each discrete state leads to the possibility of
additional interference condition, so that the presence of several
defects may show a variety of interference phenomena. In order to
study these effects in details, both analytically and numerically,
we take one of the simplest implementations of the $N$-mode Fano
defect as a finite chain of defects with the nearest-neighbor
coupling between them, see Fig.~\ref{fig:fig1}(a)
\cite{trombettoni}.

We then provide with specific example from the theory of Photonic Crystal (PC) waveguides in order
to show how this model can be implemented to real physical system.

\subsection{Model for $N$-defect Fano resonances}

We start our study from the analysis of the linear transmission
when the Hamiltonian of the model can be written in the following
form
\begin{eqnarray}
\label{eq:lin_ham}
  H_L=H_0+H_F+H_{0F}\;,
\end{eqnarray}
where
\begin{eqnarray}
\label{terms}
  H_0&=&C\sum\limits_n(\phi_n\phi_{n+1}^*+c.c.)\;,\\
  H_F&=&\sum\limits_{m=1}^{N-1}(E_m|\varphi_m|^2+V_m\varphi_m\varphi_{m+1}^*+c.c.)+E_N|\varphi_N|^2\;,\nonumber\\
  H_{0F}&=&V_0\phi_0\varphi_1^*+c.c.\;,\nonumber
\end{eqnarray}
and the asterisk stands for the complex conjugation. The model
with the Hamiltonian (\ref{eq:lin_ham}), (\ref{terms}) describes
the interaction of two subsystems coupled locally to each other at
a single site [see Fig.~\ref{fig:fig1}(a)]. One subsystem is an
infinite homogeneous array of equivalent elements described by the
wavefunctions $\phi_n$ with the strength of the nearest-neighbor
interaction characterized by the parameter $C$. In this array,
waves propagate freely and they are characterized by the
dispersion relation $\omega_q=2C\cos q$. The other subsystem is a
finite inhomogeneous chain of $N$ elements described by the
wavefunctions $\varphi_n$ which acts as a complex localized defect
attached to the main array, with $E_m$ being the energy associated
with the $m$-th element. We assume that the defect sites are
coupled through the nearest-neighbor interaction with the strength
$V_m$.

Depending on the ratio of the coupling coefficients $C$ and $V_0$
this model can be directly applied to different physical systems.
For example, in quantum dots $C\ll V_0$ \cite{goeres00,bulka01,spinfilters}, 
in photonic-crystal
waveguides coupling coefficients $C$ and $V_0$ are of the same order 
\cite{shfjdj02,shf02,mfyshfms03,aemysk05}, 
for scattering by time-periodic
and spatially localized states (Discrete Breathers) even the case $C < V_0$
might be possible \cite{sfaemvfmvf03}. 
Therefore, we will not concentrate on a particular
type of the physical system and will present generic results.

From the Hamiltonian (\ref{eq:lin_ham}), (\ref{terms}) we can derive
the equations of motion in the frequency domain,
\begin{eqnarray}
\label{eq:lin_equans}
  \omega\phi_n&=&C(\phi_{n-1}+\phi_{n+1})+V_0\varphi_1\delta_{n,0}\;,\nonumber\\
  \omega\varphi_1&=&E_1\varphi_1+V_0\phi_0+V_1\varphi_2\;,\nonumber\\
  \omega\varphi_2&=&E_2\varphi_2+V_1\varphi_1+V_2\varphi_3\;,\\
   &\vdots&\nonumber\\
   \omega\varphi_N&=&E_N\varphi_N+V_{N-1}\varphi_{N-1}\;.\nonumber
\end{eqnarray}
and obtain a simple recurrence relation for $\varphi_m$,
\begin{eqnarray}
\label{eq:recurrent1}
  \varphi_{N-l}=\frac{V_{N-l-1}\rho_{l}(\omega)}{\rho_{l+1}(\omega)}\varphi_{N-l-1}\;,
\end{eqnarray}
where
\begin{eqnarray}
\label{eq:poly}
       \rho_l(\omega)=\det[\omega I-H^l_F]
 \end{eqnarray}
is a characteristic polynomial of the subsystem of $l$ sites only
(calculated from the site $N-l+1$ to the site $N$)
\begin{eqnarray}
\label{eq:subsys_l}
 H^l_F=\left(
      \begin{array}{ccccc}
        E_N&V_{N-1}&0&\cdots&0\\
        V_{N-1}&E_{N-1}&V_{N-2}&\cdots&0\\
        \vdots&&\ddots&&\vdots\\
        0&\cdots&0&V_{N-l+1}&E_{N-l+1}
       \end{array}
     \right)
\end{eqnarray}
and $I$ is the identity matrix $l\times l$.

The recurrence relation (\ref{eq:recurrent1}) is valid for
$l=0,\dots,N-2$ and, for simplicity, we assume that
$\rho_0(\omega)\equiv1$. By writing the relation
(\ref{eq:recurrent1}) for $\varphi_2$,
\[
\varphi_2=\frac{V_1\rho_{N-2}(\omega)}{\rho_{N-1}(\omega)}\varphi_1
\]
and substituting it into the second equation of
(\ref{eq:lin_equans}), we extend the recurrence relation
(\ref{eq:recurrent1}) to the following form
\begin{eqnarray}
\label{eq:recurrent2}
  \varphi_{1}=\frac{V_{0}\rho_{N-1}(\omega)}{\rho_{N}(\omega)}\phi_{0}\;,
\end{eqnarray}
which allows us to eliminate additional degrees of freedom form
the system (\ref{eq:lin_equans}) and obtain a system of equations
for describing the wave propagation in the main array only
\begin{eqnarray}
\label{eq:lin_sys2}
  \omega\phi_n&=&C(\phi_{n-1}+\phi_{n+1})+\frac{\rho_{N-1}(\omega)}{\rho_{N}(\omega)}V_0^2\phi_{0}\delta_{n,0}\;
\end{eqnarray}
with an effective localized defect. This defect acts as a
scattering potential whose strength depends on the frequency of
the incoming wave (\ref{eq:lin_sys2}). The function
$\rho_l(\omega)$ vanishes when the frequency $\omega$ coincides
with one of the eigenfrequencies of the corresponding  subsystem
of $l$ sites described by Eq.~(\ref{eq:subsys_l}). Therefore, our
first important result is that, in general, there exist $N-1$
frequencies where the strength of the defect vanishes and the
induced scattering potential becomes transparent, and $N$
frequencies where the strength of the defect will become infinite
making the scattering potential opaque.

\subsection{Examples from the photonic crystal theory}

\begin{figure}[htb]
\centerline{
\includegraphics[width=80mm]{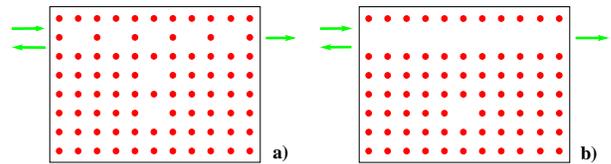}}
\caption{\label{fig:fig2}(Color online) Schematic view of two photonic crystal waveguide configurations
with side-coupled defects. The waveguides and defects are
constructed by removing some rods in a periodic structure.
These two examples show that coupling between defect rods
in straight waveguide and coupling to side-coupled defects can be easily tuned.
}%
\end{figure}

One of the important physical systems, where the model described above
can be applied, is photonic crystal waveguides. 
Some examples are shown in Fig. \ref{fig:fig2}.

Photonic crystals are artificial dielectric structures with
a periodic modulation in the refractive index that create regions
of forbidden frequencies known as photonic
band gaps ~\cite{jdjprvsf97}.
Due to the small period of the modulation  of the refractive index $\approx500\,\mathrm{ nm}$,
photonic crystals are known as nano-devices, which allow to guide light by varying the waveguide
configuration.
These waveguides are usually constructed by introducing defects in a periodic structure.

Below we demonstrate
that transmission of electromagnetic waves through 
photonic crystal waveguides can be described by a simple discrete
model, which is similar to Eq. (\ref{eq:lin_equans}).
We consider a
two-dimensional photonic crystal created by a square lattice (with
the period $a$) of dielectric rods in air.
We study in-plane light propagation in
this photonic lattice described by the electric field
$E(\mathbf{x},t)=\exp(-i\omega t)E(\mathbf{x}|\omega)$ polarized
parallel to the rods, and reduce the Maxwell's equations to the
scalar eigenvalue problem
\begin{eqnarray}
\label{electric_field}
\left[\nabla^2+\left( \frac{\omega}{c}\right)^2\epsilon(\mathbf{x})\right]E(\mathbf{x}|\omega)=0\;.
\end{eqnarray}

A waveguide is created by replacing some of the lattice rods by the
defect rods with the radius $r_d$, or simply by removing some rods
of the lattice.
To describe the structure with defects, we
decompose the permittivity function $\epsilon(\mathbf{x})$ into a
sum of the periodic part and the defect-induced contribution,
$\epsilon(\mathbf{x})=\epsilon_p(\mathbf{x})+\delta\epsilon(\mathbf{x})$,
and rewrite Eq.~(\ref{electric_field}) in the integral
form~\cite{sfmyskras00},
\begin{eqnarray}
\label{integral}
E(\mathbf{x}|\omega)=\left( \frac{\omega}{c}\right)^2\int d^2\mathbf{y}
G(\mathbf{x,y}|\omega)\delta\epsilon(\mathbf{y})E(\mathbf{y}|\omega)\;,
\end{eqnarray}
%
%

%
where $G(\mathbf{x,y}|\omega)$ is the standard Green's function.
If the radius of the defect rod $r_d$ is sufficiently small, the
electric field $E(\mathbf{x}|\omega)$ inside the rod is almost
constant, and the integral (\ref{integral}) can be easily
evaluated. This allows us to derive a set of discrete equations
for the electric field
\begin{eqnarray}
\label{discrete_model}
E_{n,m}=\sum\limits_{k,l}J_{n-k,m-l}(\omega)\delta\epsilon_{k,l}E_{k,l}\;
\end{eqnarray}
where
\begin{eqnarray}
J_{n,m}(\omega)=\left(\frac{\omega}{c}\right)^2\int\limits_{r_d} d^2\mathbf{y}
G(\mathbf{x}_n,\mathbf{x}_m+\mathbf{y}|\omega)
\end{eqnarray}
are the frequency-dependent effective coupling coefficients and
\begin{eqnarray}
\delta\epsilon_{n,m}=\epsilon_{n,m}-\epsilon_{\rm rod}\;,
\end{eqnarray}
are the defect-induced changes of the lattice dielectric function,
where $\epsilon_{n,m}$ is the dielectric constant of the defect
rod located at the site $(n,m)$.

In general, the effective coupling coefficients
$|J_{n,m}(\omega)|$ decay slow in space \cite{sfmysk02}.
This slow decay introduces effective 
long-range interaction between different sites of the waveguide.
In reality, we define a finite distance $L$ of this interaction 
by assuming that all coupling coefficients
with the numbers $|n-k|>L$ and $|m-l|>L$ vanish. As demonstrated
in Ref.~\cite{sfmysk02}, the case $L=6$ gives already an excellent
agreement with the results of the finite-difference time-domain
numerical simulations. 
But for many cases even the nearest-neighbor interaction approximation ($L=1$), shows a 
good agreement with exact results. 
In the case of side-coupled defects to the straight waveguide (see Fig. \ref{fig:fig2}(a)),
the set of equations (\ref{discrete_model}) reduces exactly to (\ref{eq:lin_sys2}). 
This drastic simplification of the original problem allows us to study the system
analytically and analyze many interesting effects such as resonant light scattering. 
By taking into account a larger number of interaction terms will just renormalize the effect.

\subsection{Transmission coefficient}

To calculate the transmission coefficient for the system (\ref{eq:lin_sys2}), we use the transfer
matrix connecting the left and right boundaries \cite{ptblbh99}
\begin{eqnarray}
\label{eq:tr_matrix}
  \left(\begin{array}{c}\phi_{L}\\ \phi_{L+1}\end{array}\right)=%
   M \left(\begin{array}{c}\phi_{-L}\\ \phi_{-L-1}\end{array}\right)\;,
\end{eqnarray}
where $2L$ is a characteristical width of the scattering potential.

By using the scattering  boundary conditions
\begin{eqnarray}
\label{eq:scatt_bound}
\phi_n = \left\lbrace %
 \begin{array}{lc}
    I_0\mathrm{e}^{iqn}+r\mathrm{e}^{-iqn},& n\ll-L,\\
    t\mathrm{e}^{iqn}, & n\gg L,
   \end{array}
   \right.
\end{eqnarray}
the transmission coefficient $T=|t^2|/|I_0|^2$ can be presented in
the following form
\begin{eqnarray}\label{eq:transmission}
    T=\frac{4\sin^2q}{\left|M_{11}e^{-iq}+M_{12}-M_{21}-M_{22}e^{iq}\right|^2}.
\end{eqnarray}
For the $\delta$-like scattering potential (\ref{eq:lin_sys2}) the
transfer matrix (\ref{eq:tr_matrix}) takes a very simple form
\begin{eqnarray}
\label{eq:M_F}
  M_F=\left(
    \begin{array}{cc}
      a&-1\\
      1&0
    \end{array}
    \right),
\end{eqnarray}
where
\begin{equation}
 a=2\cos q-\frac{V_0^2}{C}\frac{\rho_{N-1}(\omega_q)}{\rho_N(\omega_q)}
\end{equation}
and it defines the following transmission coefficient
\begin{equation}
\label{eq:Ntransm}
T=\frac{\alpha_q^2}{\alpha_q^2+1},
\end{equation}
where
\begin{equation}
\label{eq:Ntransm2}
\alpha_q=\frac{c_q\rho_N(\omega_q)}{V_0^2\rho_{N-1}(\omega_q)},
\;\; c_q=2C\sin q.
\end{equation}

From the result (\ref{eq:Ntransm}), (\ref{eq:Ntransm2}) it follows
that, in general, there exist $N-1$ frequencies for the perfect
transmissions (when $\alpha_q=\pm\infty$) and $N$ frequencies for
the perfect reflections (when $\alpha_q=0$). Due to the specific
structure of our system, the perfect transmissions are surrounded by
the perfect reflections. The transmission coefficient
(\ref{eq:Ntransm}), (\ref{eq:Ntransm2}) is written the form similar
to that of the Fano formula, and it allows us to associate the
resonances with the Fano resonance.

\begin{figure}[htb]
\centerline{
\includegraphics[width=80mm]{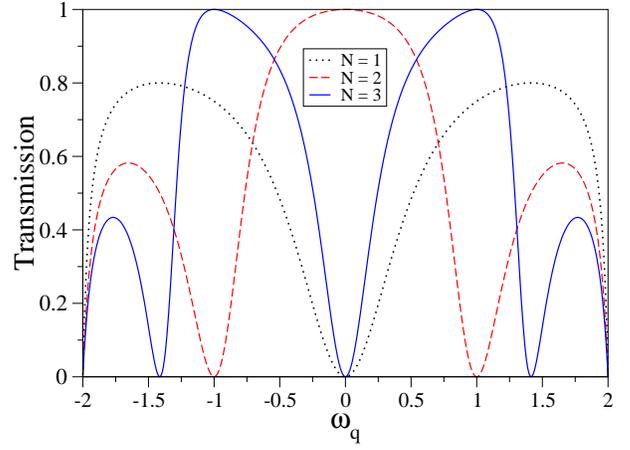}}
\caption{\label{fig:fig3}(Color online) Transmission coefficient of the
$N$-level Fano defect with the degenerated energies $E_m=E$ for
$N=1, 2, 3$. Other parameters are $C=1$, $V_m=1$, and $E=0$.
}%
\end{figure}

The case of local coupling considered above is quite special, and it
may be hard to realize in a real physical system. Nevertheless, it
allows us to demonstrate the entire physical phenomenon and
associate the resonant reflection and transmissions with the
excitation of particular groups of the defects. In particular, by
analyzing these results we make the following statements about the
nature of the Fano resonances. (i) The frequencies of {\em perfect
reflections} occur at the eigenmode frequencies of the complex
$N$-site Fano defect. In order to find these frequencies, we should
cut the coupling between the main array and a finite subsystem of
defects and calculate a discrete spectrum of oscillatory frequencies
of the isolated complex defect. This result agrees with the earlier
results obtained for other types of time-periodic and
spatially-localized scattering potentials~\cite{sfaemvfmvf03}. (ii)
The frequencies of {\em perfect transmissions} can be calculated by
eliminating one degree of freedom from the a cluster of defects, as
indicated in Fig.~\ref{fig:fig1}(a) and calculating the oscillatory
eigenfrequencies of the remaining chain. Because, both the perfect
reflections and perfect transmissions excite some eigenstates of the
complex defect they correspond to a resonant scattering.

As an example, we consider a simple case when the complex Fano
defect consists of a homogeneous chain of defects, and the energy
of all sites and couplings between them are constant, $E_m=E$ and
$V_m=V$. In this case, the function $\rho_l(\omega)$ vanishes at
\begin{eqnarray}
  \omega_m=E+2V\cos\left(\frac{m\pi}{l+1}\right)\;,\;\;\;m=1,\ldots,l\;,
\end{eqnarray}
which leads to the corresponding resonances as shown in
Fig.~\ref{fig:fig3}. This example shows a very interesting property
of the Fano resonances. By adding or removing one additional defect
we can change the transmission from zero to one for some particular
frequencies. Namely, each perfect reflection of the $l$-level Fano
defect with $T_l(\omega_m)=0$ becomes a perfect transmission with
$T_{l+1}(\omega_m)=1$ after adding one more defect to the chain.

\subsection{Additional defects in the main array}

Now we study another important case that allows an effective
engineering of the Fano resonance transmission. In particular, we
study the effect of an additional defect placed in the main array.
In this case, the effective equation becomes
\begin{eqnarray}
\label{eq:lin_sysD}
  \omega\phi_n=C(\phi_{n-1}&+&\phi_{n+1})+E_0\phi_{l1}\delta_{n,l_1}+\nonumber\\
  &+&\frac{\rho_{N-1}(\omega)}{\rho_{N}(\omega)}V_0^2\phi_{0}\delta_{n,0}\;,
\end{eqnarray}
where we assume that the $\delta$-like defect of the strength $E_0$
is located at the site $n=l_1$ and the complex $N$-level Fano defect
remains at the site $n=0$ [see Fig.~\ref{fig:fig1}(b)].

Without losing the generality, we assume that $l_1\ge0$, and present
the transfer matrix $M$ in the following form
\begin{eqnarray}
\label{eq:tr_matrixD}
        M&=&M_{\delta}M_0^mM_F,\\
   M_0 = \left(
      \begin{array}{cc}
        b&-1\\
        1&0
       \end{array}
     \right),&&
    M_{\delta} = \left(
      \begin{array}{cc}
        c&-1\\
        1&0
       \end{array}
     \right),\nonumber\\
  b=2\cos q, && c=b-E_0/C, \nonumber
\end{eqnarray}
and $m=l_1-1$. By using the eigenvalue expansion, we can show that
\begin{eqnarray}
\label{eq:tr_matrix_M0}
        M_0^m = \frac{1}{\sin q}\left(
      \begin{array}{cc}
        \sin[(m+1)q]&-\sin[mq]\\
        \sin[mq]&-\sin[(m-1)q]
       \end{array}
     \right),
\end{eqnarray}
and calculate analytically the transfer matrix $M$ and the
transmission coefficient (\ref{eq:transmission}). Again, we
observe that there exist $N$ perfect reflections at the
frequencies of the eigenstates of the $N$-level Fano defect, as
discussed above. However, the condition for perfect transmissions
becomes more complicated.

\begin{figure}[htb]
\centerline{
\includegraphics[width=80mm]{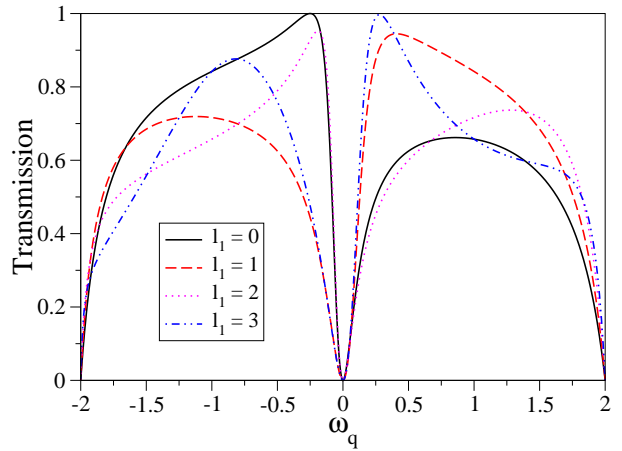}}
\caption{\label{fig:fig4}(Color online) Transmission coefficient of combined
Fano and $\delta$-like defects for different distances between
them. Other parameters are $C=1$, $V_0=0.5$, $E_0=1$, and $E_1=0$.
}%
\end{figure}

First, we consider a single Fano defect ($N=1$) and study how the
transmission depends on the distance $l_1$ between two defects.
The transfer matrix (\ref{eq:tr_matrixD}) possesses the same singularity 
as the transfer matrix of the single Fano defect (\ref{eq:M_F}), which
leads to the perfect reflection at the same resonant frequency 
(see Fig.\ref{fig:fig3}).
Figure~\ref{fig:fig4} shows clearly that, in addition to the
perfect reflection, there exists a resonant
transmission due to the presence of the defect in the main array.
The transmission function becomes {\em  asymmetric}, and it can be
described by the generalized Fano formula. We notice here a quite
interesting behavior: the transmission coefficient alternates 'the
sing of asymmetry', i.e. $\omega_{T_{\rm max}}<\omega_{T_{\rm
min}}$, for even values of $l_1$, and $\omega_{T_{\rm
max}}>\omega_{T_{\rm min}}$, for odd values of $l_1$. Moreover,
the maximum of the transmission does not reach one in some cases
(see Fig.~\ref{fig:fig4}). One of the possible explanations of
this effect is that a plane wave accumulates an additional phase
shift propagating between the two defects, which leads to the {\em
effective decoherence} and, as a result, incomplete interference.
But this effect does not alter the perfect reflection, and this
reveals the principal difference between the resonant reflections
and the resonant transmissions associated with the Fano resonance.

In order to show more clearly the difference between the resonant
reflection and the resonant transmission at the Fano resonance, we
consider the case when the $\delta$-defect in the main array and
the $N$-level Fano defect are  coupled directly, i.e. when
$l_1=0$, which leads to a generalized point defect in the system
(\ref{eq:lin_sysD}). In this case, the transmission coefficient
takes the form
\begin{eqnarray}
T = \frac{4C^2\rho_{N}(\omega)^2\sin^2
q}{\left[\rho_{N-1}(\omega)V_0^2+E_0\rho_{N}(\omega)\right]^2
+4C^2\rho_{N}(\omega)^2\sin^2 q},
\end{eqnarray}
and the condition for perfect transmissions is
\begin{eqnarray}
\label{polinomial}
  \rho_{N-1}(\omega)V_0^2+E_0\rho_{N}(\omega)=0.
\end{eqnarray}
Equation (\ref{polinomial}) is a polynomial of the order $N$. This
means that, in general, there exists a possibility for $N$
frequencies of the perfect transmissions. Such frequencies {\em do
not coincide} with the oscillatory eigenfrequencies of the
$N-1$-defect system, as discussed above (see Fig.~\ref{fig:fig1}),
or $N+1$-defect system (i.e. $N$-level Fano defect plus a single
$\delta$-like defect), as one can expect. Therefore, this means
that perfect transmissions do not necessarily correspond to a
resonant behavior. In this case, the dependence of the
transmission coefficient is flat and, therefore, the scattering
potential is almost transparent. The situation changes
dramatically when the frequency of perfect transmission is located
very close to that of perfect reflection, i.e.
$\omega_{T=1}\approx\omega_{T=0}$. In the latter case, the perfect
transmission becomes resonant because it corresponds to the
excitation of one of the eigenmodes of the Fano defect, which is
responsible for the resonant suppression of the transmission. This
physical picture explains how the Fano defect can generate, almost
for the same frequency, both resonant constructive and destructive
interferences creating a sharp asymmetric profile of the
transmission curves. We would like to emphasize here again that
the main feature of the Fano resonance is a resonant reflection
rather than transmission, and the associated perfect transmission
itself could be or could not be resonant.

\begin{figure}[htb]
\centerline{
\includegraphics[width=75mm]{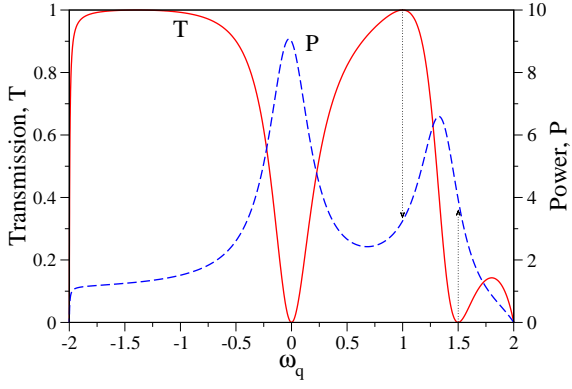}}
\caption{\label{fig:fig5}(Color online) Transmission through the $N$-level Fano
defect with $N=3$ and a single $\delta$-like defect in the main
array for $E_m=\Delta+mE$ ($m=\bar{0:N}$) and the parameters
$C=1$, $V_m=1$, $\Delta=0.5$, and $E=0.5$. The effective norm
(\ref{eq:norm2}) of the Fano defect is also shown to characterize
the resonant excitation.
}%
\end{figure}

Based on the analysis presented above, we can characterize the
resonant scattering qualitatively by monitoring the strength of
the excitation of the Fano defect. For that purpose, we introduce
an effective power of the $N$-level Fano defect as the following
norm,
\begin{eqnarray}
\label{eq:norm}
    \Psi=\sum\limits_{m=1}^N|\varphi_m|^2\;,
\end{eqnarray}
where, for simplicity, we assume $V_m\equiv V$. By using the
recurrence relation (\ref{eq:recurrent1}), we express all
amplitudes $\varphi_m$ in terms of $\varphi_1$,
\begin{eqnarray}
    |\varphi_m|^2=\frac{V^{2m-2}\rho_{N-m}^2(\omega)}{\rho_{N-1}^2(\omega)}|\varphi_1|^2
\end{eqnarray}
We use the relation (\ref{eq:recurrent2}) and employ the fact that
the norm at site $n=0$ for a $\delta$-like scattering potentials
is proportional to the transmission coefficient
$|\phi_0|^2=T|I_0|^2$. This allows us to write the norm
(\ref{eq:norm}) in the following form
\begin{eqnarray}
\label{eq:norm2}
    \Psi=\frac{T|I_0|^2}{\rho_N^2(\omega)}\sum\limits_{m=1}^NV^{2m}\rho_{N-m}^2(\omega).
    \end{eqnarray}

\begin{figure}[htb]
\centerline{
\includegraphics[width=75mm]{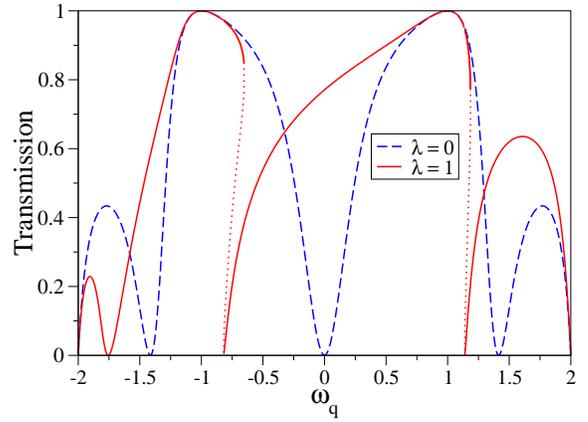}}
\caption{\label{fig:fig6}(Color online) Nonlinear transmission through the $N=3$
defect chain with $E_m=E$ coupled to the main array. Other
parameters are $C=1$, $V_m=1$, $E=0$, $\lambda=1$, and $I=1$. For
these parameters the shifts of the perfect reflections are large
enough to show the presence of bistability. The region of the
bistable transmission is indicated by the dotted line. For
reference, we show the linear transmission at $\lambda=0$.
}%
\end{figure}

As an example, we consider the transmission trough the
$(N+1)$-level $\delta$-defect, which consists of a single
$\delta$-like defect in the main array and $N$-level Fano defect,
as shown in Fig.~\ref{fig:fig1}(b). We assume the equidistant
energy levels of the defects, $E_m=\Delta+mE$, $m=0,\ldots,3$, and
constant coupling, $V_m=V$. We observe that inside the
transmission spectrum there exist two frequencies of the prefect
transmission and two frequencies of the perfect reflection. One
perfect transmission is resonant (at $\omega\approx1$) and the
other one is not (at $\omega\approx-1.5$). In Fig.\ref{fig:fig5}
we plot also the norm of the Fano defect defined by
Eq.~(\ref{eq:norm2}). It shows that the nonresonant perfect
transmission does not excite the Fano defect, and it is related to
a simple constructive interference. The frequency of the other
perfect transmission is located close to that of the perfect
reflection, and it corresponds to a strong excitation of the Fano
defect making the perfect transmission resonant.  In this case,
the strengths of the Fano defect excitation at the perfect
transmission and perfect reflection almost coincide.

\section{Nonlinear transmission}

The analytical and numerical results presented above show that the
resonant reflections associated with the Fano resonances are
robust in the regime of a local coupling. Such reflections are
observed when the frequency of the incoming wave coincides with
one of the frequencies of the oscillatory eigenmodes of the
attached defect chain. As a result, the defects become highly
excited at the frequency of the resonant reflection. Such a
specific resonant reflection can be tuned externally, and below we
discuss how the presence of nonlinear defects in the discrete
network may be employed to manage and tune the response of the
Fano resonances.

We consider the general case of the $N$-defect Fano resonance and
assume that one of the defects possesses a Kerr-type nonlinear
response that can contribute as an additional nonlinear term into
the system Hamiltonian,
\begin{eqnarray}
  H_{NL}=H_L+\lambda|\varphi_1|^4\;.
\end{eqnarray}
For definiteness, we choose the first defect $\varphi_1$ as
nonlinear due to its specific role in the transmission properties
and resonant reflections. Similar to the earlier studies, we
expect that the presence of such a nonlinear defect will shift the
positions of the perfect reflection depending on the intensity of
the incoming wave~\cite{aemsfmsfysk05}, while the perfect
transmissions will be unchanged or modified only slightly. This
feature would allow us to achieve a simple tuning of the width of
the asymmetric Fano resonance by changing the intensity of the the
incoming wave.

In the presence of the nonlinear defect, the equations of the
motion can be written as follows
\begin{eqnarray}
\omega\phi_n&=&C(\phi_{n-1}+\phi_{n+1})+V_0\varphi_1\delta_{n,0}\;,\nonumber\\
  \frac{\rho_N(\omega)}{\rho_{N-1}(\omega)}\varphi_1&=&\lambda|\varphi_1|^2\varphi_1+V_0\phi_0\;,
\end{eqnarray}
and these equations are similar to the equations of the recent
paper~\cite{aemsfmsfysk05} for describing the resonant
transmission of a single nonlinear defect at the Fano resonance.

Using the approach developed earlier in Ref.~\cite{aemsfmsfysk05},
we obtain the result for the transmission coefficient
 \begin{eqnarray}
T=\frac{x^2}{x^2+1}\;,
\end{eqnarray}
where $x$ is a real solution of the cubic equation
\begin{eqnarray}
\label{eq:cubic}
(x^2+1)(x-\alpha_q)-\gamma_q =0,
\end{eqnarray}
with the parameter $\gamma_q=\lambda c_q^3|I|^2/V_0^4$. The
perfect reflection ($T=0$) takes place when there exists zero
solution $x=0$ of the cubic equation (\ref{eq:cubic}), and this
becomes possible when $\alpha_q=-\gamma_q$. As a result, the
presence of nonlinearity leads to a shift of the position of the
perfect reflection in comparison with the case of the linear
transmission at $\alpha_q=0$ described by Eq.~(\ref{eq:Ntransm}).

When the system allows for several perfect reflections inside the
transmission spectrum, the shift depends on the position and the
frequencies of the resonant reflections are shifted nonuniform,
see an example in Fig.~\ref{fig:fig6}. In contrast, the
frequencies of the perfect transmission ($T=1$) remains unchanged
since they correspond to the conditions $x=\infty$ or
$\alpha_q=\infty$. As was already mentioned in
Ref.~\cite{aemsfmsfysk05}, nonlinear transmission may become
unstable and bistable when $|\alpha_q|^2>3$. Therefore, near the
scattering resonances bistable transmission can be observed under
some conditions, as shown in the example presented in
 Fig.~\ref{fig:fig6}.

\section{Conclusions}

We have suggested an effective way to engineer the resonant wave
transmission and reflection in discrete networks through the
concept of the Fano resonance management. In particular, we have
analyzed the transmission properties of a linear array of
interacting elements coupled to a chain of $N$ side-coupled
defects and found exact analytical solutions for the transmission
coefficient and the conditions for the perfect reflections and
transmissions due to either destructive or constructive
interferences. We have demonstrated  that the nature of these
reflections and transmissions can be associated with the familiar
concept of Fano resonances, and we have formulated the basic
principles of the resonant scattering management by tuning the
Fano resonances. In addition, we have presented an example of a
nonlinearity-tunable Fano resonance when one defect of the network
possesses a nonlinear Kerr-like response. We believe our findings
and the basic physical concepts of the $N$-defect transmission and
reflection will be useful for many various systems where the
resonant transmission can be characterized and described in terms
of the Fano resonances.

\section*{Acknowledgments}

We thank S.~Flach and S.~Mingaleev for useful collaborations. This
work was partially supported by the Australian Research Council
through the Centre of Excellence Program. Nonlinear Physics Centre
is a member of the Centre for Ultra-high bandwidth Devices for
Optical Systems (CUDOS).

\end{document}